\documentclass[prd,aps,showpacs,nofootinbib]{revtex4}%
\usepackage{graphicx}
\usepackage{amsmath}
\usepackage{amsfonts}
\usepackage{amssymb}%
\newcommand{\be}{\begin{equation}}
\newcommand{\ee}{\end{equation}}
\newcommand{\bq}{\begin{eqnarray}}
\newcommand{\eq}{\end{eqnarray}}
\begin{document}
\title{\textbf{Noncommutative Geometry and Cosmology}}
\author{G. D. Barbosa}
\email{gbarbosa@cbpf.br}
\author{N. Pinto-Neto}
\email{nelsonpn@cbpf.br}
\affiliation{Centro Brasileiro de Pesquisas F\'{\i}sicas, CBPF, Rua Dr. Xavier Sigaud 150,
22290-180, Rio de Janeiro, Brazil }

\begin{abstract}
We study some consequences of\ noncommutativity to homogeneous cosmologies by
introducing a deformation of the commutation relation between the
minisuperspace variables. The investigation is carried out for the
Kantowski-Sachs model by means of\ a comparative study of the universe
evolution in four different scenarios: the classical commutative, classical
noncommutative, quantum commutative, and quantum noncommutative. The
comparison is rendered transparent by the use of the Bohmian formalism of
quantum trajectories. As a result of our analysis, we found that
noncommutativity can modify significantly the universe evolution, but cannot
alter its singular behavior in the classical context. Quantum effects, on the
other hand, can originate non-singular periodic universes in both commutative
and noncommutative cases. The quantum noncommutative model is shown to present
interesting properties, as the capability to give rise to non-trivial dynamics
in situations where its commutative counterpart is necessarily static.

\end{abstract}
\maketitle

\section{Introduction}

Recently, there has been a great amount of work devoted to noncommutative
theories (see, e.g., \cite{1,2}). The boom of interest in noncommutativity of
the canonical type was triggered by works establishing its connection with
string and M-theory \cite{3}, although previous investigations in the context
of semiclassical gravity \cite{4}\ have pointed out the relevance of
noncommutative field theories. In addition to its relevance to string theory,
the study of noncommutative theories is justified in its own by the
opportunity it gives us to deal with interesting\ properties, such\ as the
IR/UV mixing and nonlocality \cite{5}, Lorentz violation \cite{6}, and new
physics at very short scale distances \cite{1,2,7}.

In the latest two years, several investigations have been carried out to
clarify the possible role of noncommutativity in the cosmological scenario in
a great variety of contexts. Among them, we quote the Newtonian cosmology
\cite{11}, cosmological perturbation theory and noncommutative inflationary
cosmology \cite{12}, noncommutative gravity \cite{14}, and quantum cosmology
\cite{15}. The latter, in particular, provides an interesting arena for
speculation on the possible connection between noncommutativity and quantum
gravity. The common claim that noncommutativity leads to fuzzyness renders
obscure the application of noncommutative ideas to the description of a
quantum universe, which, according to the Copenhagen interpretation of quantum
theory, has no objective reality. Indeed, the inadequacy of the Copenhagen
interpretation for quantum cosmology has been stressed long time ago by
several physicists, as Everett \cite{15.5}, Feynman\ \cite{16}, and Bell
\cite{16.5}. Recently, 't Hooft \cite{17} has argued that a reconsideration of
hidden-variables theories is in turn necessary to account for the difficulties
that appear in the unification of General Relativity with Quantum Theory.

Perhaps the great riddle of quantum gravity is the comprehension of the
behavior of spacetime (if this concept has a meaning) at the Planck scale. At
very early times, when the universe was small and hot, even when its
characteristic length scale\ was larger than the Planckian one,
noncommutativity may have played a relevant role in its evolution. The aim of
this work is to exploit this possibility by\ carrying out a comparative study
of the universe evolution in four different scenarios: classical commutative,
classical noncommutative, quantum commutative, and quantum\ noncommutative. As
our object of analysis,\ we chose the Kantowski-Sachs universe (see, e.g.,
\cite{18,18.1,18.2,18.3,18.4,18.5}). A noncommutative version of the
Kantowski-Sachs universe was previously considered in \cite{15}, where\ a
Moyal deformation of the Wheeler-DeWitt equation in the minisuperspace
approximation was introduced. In the present work, we explore another
possibility that can be present in noncommutative quantum cosmology using the
same framework pioneered in \cite{15}. The noncommutative geometry considered
here is a property of the minisuperspace observables which refer to the
physical metric. This, as will be detailed later, entails an identification of
the universe degrees of freedom as a set of variables that differs from the
one adopted in\ \cite{15}. As quoted in that reference, what renders the
Kantowski-Sachs model attractive for investigation in the noncommutative
context is the opportunity its noncommutative quantum version provides us to
deal with nonperturbative effects of noncommutative geometry and\ quantum
gravity comprised in the same model. This is assured by the analytic solutions
admitted by\ the noncommutative Wheeler-DeWitt equation.

As an interpretation for quantum theory, we are adopting Bohm's ontological
one \cite{24,24.5,19,20}. Such an interpretation has also been employed in
other works on quantum cosmology and\ quantum gravity\ (see, e.g.,
\cite{20.5,21,21.5,21.7}). In this work, with the aid of the Bohmian
minisuperspace trajectories,\ we will show how it is possible to conceive a
\textquotedblleft noncommutative quantum universe\textquotedblright\ with
minisuperpace operators satisfying a noncommutative algebra. We shall study
the quantum universe as a well defined entity, without appeal to any external
observer, which would find it \textquotedblleft fuzzy\textquotedblright\ in a
supposed measurement process. Independent of the orientation with respect to
the foundations of quantum theory one may have, the Bohmian formalism of
trajectories can always be adopted as a useful tool in providing an intuitive
interpretation for the\ quantum phenomena. Indeed, the interest in the Bohmian
approach is growing in a broad community (see, e.g., \cite{22}).

The paper is organized as follows. In section 2 we summarize the essential
aspects of canonical quantum gravity and Bohmian quantum physics necessary for
the remaining sections. Sections 3, 4 and 5 are devoted to a comparative study
of the commutative and noncommutative \ classical Kantowski-Sachs universes.
In section 6 the commutative quantum version of the model is studied in the
language of Bohmian trajectories. An extension of the Bohmian formalism is
proposed and applied to the noncommutative quantum Kantowski-Sachs universe in
section 7. In Section 8 we end up with a general discussion and summary of the
main results.

\section{Quantum Gravity and the Bohmian Formalism}

Before introducing noncommutativity in the cosmological scenario, it is
interesting to comment some aspects of the standard Hamiltonian and Bohmian
quantum gravity formalisms.

The Hamiltonian of General Relativity is usually expressed in the ADM
formulation \cite{22.5}. In this formalism, the line element is written as
\begin{equation}
ds^{2}=\left(  N_{i}N^{i}-N^{2}\right)  dt^{2}+2N_{i}dx^{i}dt+h_{ij}%
dx^{i}dx^{j}, \label{2}%
\end{equation}
where $N$ represents the lapse function, $N^{i}$ is the shift vector, and
$h_{ij}$ is the three-metric of a three-surface embedded in spacetime. The
dynamics of the spacetime is described in terms of the evolution of $h_{ij}$
in superspace, the space of all three-geometries.

The Hamiltonian of General Relativity without matter is\footnote{We shall
restrict all of our considerations to pure gravity, since this is the case of
interest in this work.}
\begin{equation}
H=\int d^{3}x\left(  N\mathcal{H}+N_{j}\mathcal{H}^{j}\right)  , \label{3}%
\end{equation}
where
\begin{equation}
\mathcal{H}=G_{ijkl}\Pi^{ij}\Pi^{kl}-h^{1/2}R^{(3)}\text{, \ }\mathcal{H}%
^{j}=2D_{i}\Pi^{ij}. \label{4}%
\end{equation}
Units are chosen such that $\hbar=c=$ $16\pi G=1$. The quantity $R^{(3)}$ is
the intrinsic curvature of the spacelike hypersurfaces, $D_{i}$ is the
covariant derivative with respect to $h_{ij},$ and $h$ is the determinant of
$h_{ij}$.\ The momentum $\Pi_{ij}$ canonically conjugated to $h^{ij}$, and the
DeWitt metric $G_{ijkl}$ are
\begin{equation}
\Pi_{ij}=-h^{1/2}\left(  K_{ij}-h_{ij}K\right)  ,\text{\ } \label{5}%
\end{equation}%
\begin{equation}
G_{ijkl}=\frac{1}{2}h^{-1/2}\left(  h_{ik}h_{jl}+h_{il}h_{jk}-h_{ij}%
h_{kl}\right)  , \label{6}%
\end{equation}
where $K_{ij}=-(\partial_{t}h_{ij}-D_{i}N_{j}-D_{j}N_{i})/(2N)$ is the second
fundamental form.

As long as one follows the Dirac\ quantization procedure, the constraints of
the theory become conditions imposed on the possible states of the
wavefunctional of the universe. The super-Hamiltonian constraint
$\mathcal{H\approx}0$ yields the Wheeler-DeWitt equation \footnote{A
particular factor ordering was assumed in its derivation.}%

\begin{equation}
\left(  G^{ijkl}\frac{\delta}{\delta h^{ij}}\frac{\delta}{\delta h^{kl}%
}+h^{1/2}R^{(3)}\right)  \Psi\lbrack h^{ij}]=0, \label{9}%
\end{equation}
which determines the evolution of the wavefunctional. Up to now, the
implications of this equation to quantum cosmology are still under debate.
Among the variety of technical and conceptual problems under discussion, there
are the issue of time and the definition of probability (see \cite{23,23.5}
and references therein). One way to circumvent them is by adopting the
non-epistemological interpretation for quantum theory proposed by Bohm
\cite{24,24.5,19,20}.

In the Bohmian approach to quantum theory an ontology is given to the physical
systems (particles, fields, etc.), which evolve continuously in time obeying a
deterministic law of motion \cite{24,24.5,19,20}. When applying the Bohmian
formalism in the study of three-space geometry evolution in quantum cosmology,
we expect that the notion of space and time should have an objective meaning,
in a similar way as the notion of trajectories has in Bohmian non-relativistic
quantum mechanics \cite{20}. Indeed, this is exactly the case in Bohmian
quantum gravity, which has the evolution law for the three-space metric
$h_{ij}$\ given by \cite{20.5}%

\begin{equation}
\Pi_{ij}=-h^{1/2}\left(  K_{ij}-h_{ij}K\right)  =\operatorname{Re}\left\{
\frac{1}{\Psi^{\ast}\Psi}\left[  \Psi^{\ast}\left(  -i\frac{\delta}{\delta
h^{ij}}\right)  \Psi\right]  \right\}  =\frac{\delta S}{\delta h^{ij}},
\label{10}%
\end{equation}
where $S$ is found by writing the wavefunctional in the polar form $\Psi
=A\exp(iS)$. An intuitive picture of the deviation of the classical behavior
present in (\ref{10}) may be constructed by substituting $\Psi=A\exp(iS)$ in
(\ref{9}) and separating the real and imaginary parts. As a result, we obtain
the equations
\begin{equation}
G^{ijkl}\frac{\delta S}{\delta h^{ij}}\frac{\delta S}{\delta h^{kl}}%
-h^{1/2}R^{(3)}+Q=0, \label{12}%
\end{equation}%
\begin{equation}
G^{ijkl}\frac{\delta S}{\delta h^{ij}}\left(  A^{2}\frac{\delta S}{\delta
h^{kl}}\right)  =0, \label{13}%
\end{equation}
where
\begin{equation}
Q=-\frac{1}{A}G^{ijkl}\frac{\delta^{2}A}{\delta h^{ij}\delta h^{kl}}.
\label{14}%
\end{equation}
Expression (\ref{12}) can interpreted as a quantum Hamilton-Jacobi equation.
The quantity $Q$, absent in the classical\ Hamilton-Jacobi equation, is known
as the\ quantum potential. It is responsible for the quantum effects present
in the three-space geometry evolution. The classical limit of the theory is
found in the regime where $Q$ is negligible if compared with the other terms
present in (\ref{12}). When this is the case, the theory is clearly reduced to
classical General Relativity in the Hamilton-Jacobi formulation. Note that,
for the evolution law (\ref{10}) be consistent, the\ wavefunctional need not
necessarily be normalizable. In Bohmian quantum gravity probability is
a\ concept that is derived from the ontology of the theory.

\section{The Kantowski-Sachs Universe}

The Kantowski-Sachs universe \cite{18} is one of the most investigated
anisotropic cosmological models. Part of the interest in this universe model
is due to the wide set of analytical solutions it admits, even if particular
types of matter are coupled to gravity. In other anisotropic models, such as
the Bianchi IX, e.g., the computation of analytical solutions is a rather
complicated task \cite{18.1}. Several investigations of the\ Kantowski-Sachs
model have been carried out recently in a great variety of contexts, such as
braneworld cosmology \cite{18.2}, scalar field cosmology \cite{18.3} and
quantum cosmology \cite{18.1,18.4}. In addition to its cosmological relevance,
the Kantowski-Sachs geometry might be useful in the description of the black
holes. It has the same symmetries as the spatially homogeneous interior region
of the extended vacuum Kruskal solution that represents the late stage of
evolution of an isotropic black hole when the matter can be neglected. Indeed,
a possible connection between the Kantowski-Sachs metric with quantum black
holes and quantum wormholes has been proposed \cite{18.5}.

The Kantowski-Sachs line element is \cite{18}%

\begin{equation}
ds^{2}=-N^{2}dt^{2}+X^{2}(t)dr^{2}+Y^{2}(t)\left(  d\theta^{2}+\sin^{2}\theta
d\varphi^{2}\right)  . \label{14.5}%
\end{equation}
In the Misner parametrization, (\ref{14.5}) is written as \cite{15}
\begin{equation}
ds^{2}=-N^{2}dt^{2}+e^{2\sqrt{3}\beta}dr^{2}+e^{-2\sqrt{3}\beta}e^{-2\sqrt
{3}\Omega}\left(  d\theta^{2}+\sin^{2}\theta d\varphi^{2}\right)  . \label{15}%
\end{equation}
From (\ref{3}) and (\ref{4}), the Hamiltonian of General Relativity for this
metric is found to be
\begin{equation}
H=N\mathcal{H}=N\exp\left(  \sqrt{3}\beta+2\sqrt{3}\Omega\right)  \left[
-\frac{P_{\Omega}^{2}}{24}+\frac{P_{\beta}^{2}}{24}-2\exp\left(  -2\sqrt
{3}\Omega\right)  \right]  . \label{16}%
\end{equation}

A good characterization of the evolution of\ the spacetime metric (\ref{15})
is provided by the study of its volume expansion $\Theta\equiv V_{;\alpha
}^{\alpha}$ with respect to the commoving observer using proper time,
$V^{\alpha}=\delta_{0}^{\alpha}/N$, and\ the shear $\sigma^{2}=\sigma
^{\alpha\beta}\sigma_{\alpha\beta}/2$, where $\sigma_{\alpha\beta}=(h_{\alpha
}^{\mu}h_{\beta}^{\nu}+h_{\beta}^{\mu}h_{\alpha}^{\nu})V_{\mu;\nu}/2-\Theta
h_{\alpha\beta}/3$. The semicolon stands for four-dimensional covariant
derivative, and $h_{\alpha}^{\mu}=\delta_{\alpha}^{\mu}+V^{\mu}V_{\alpha}$ is
the projector orthogonal to the observer $V^{\alpha}$ \cite{18.7}. A
characteristic length scale $l$ can also be defined in terms of the volume
expansion through $\Theta=3\dot{l}/(lN)$. In the gauge $N=24\exp\left(
-\sqrt{3}\beta-2\sqrt{3}\Omega\right)  ,$ the volume expansion, shear, and
characteristic volume for the Kantowski-Sachs metric read%

\begin{align}
\Theta(t)  &  =\frac{1}{N}\left(  \frac{\dot{X}}{X}+2\frac{\dot{Y}}{Y}\right)
=-\frac{\sqrt{3}}{24}\left(  \dot{\beta}+2\dot{\Omega}\right)  \exp\left(
\sqrt{3}\beta+2\sqrt{3}\Omega\right)  ,\nonumber\\
\sigma(t)  &  =\frac{1}{N\sqrt{3}}\left(  \frac{\dot{X}}{X}-\frac{\dot{Y}}%
{Y}\right)  =\frac{1}{24}\left(  2\dot{\beta}+\dot{\Omega}\right)  \exp\left(
\sqrt{3}\beta+2\sqrt{3}\Omega\right)  ,\label{17}\\
l^{3}\left(  t\right)   &  =X\left(  t\right)  Y^{2}\left(  t\right)
=\exp\left(  -\sqrt{3}\beta\left(  t\right)  -2\sqrt{3}\Omega\left(  t\right)
\right)  .\nonumber
\end{align}

\section{Commutative Classical Model}

In order to distinguish individually the role of the quantum and
noncommutative effects in our noncommutative quantum universe, it is
interesting to start our study of the Kantowski-Sachs geometry\ from its
commutative classical version, which will be our reference for comparison
later. The Poisson brackets for the classical phase space variables are
\begin{equation}
\left\{  \Omega,P_{\Omega}\right\}  =1\text{, \ }\left\{  \beta,P_{\beta
}\right\}  =1\text{,\ \ }\left\{  P_{\Omega},P_{\beta}\right\}  =0\text{,
\ }\left\{  \Omega,\beta\right\}  =0. \label{18}%
\end{equation}
For the metric (\ref{15}),\ the super-Hamiltonian constraint $\mathcal{H}%
\approx0$\ is reduced to
\begin{equation}
\mathcal{H}=\xi h\approx0, \label{19}%
\end{equation}
where
\begin{equation}
\xi=\frac{1}{24}\exp\left(  \sqrt{3}\beta+2\sqrt{3}\Omega\right)
\text{,\ \ }h=-P_{\Omega}^{2}+P_{\beta}^{2}-48\exp\left(  -2\sqrt{3}%
\Omega\right)  \approx0. \label{20}%
\end{equation}

The classical equations of motion for the phase space variables $\Omega$,
$P_{\Omega},$ $\beta$ and $P_{\beta}$ are
\begin{align}
\dot{\Omega}  &  =N\left\{  \Omega,\mathcal{H}\right\}  =-2P_{\Omega
},\nonumber\\
\dot{P}_{\Omega}  &  =N\left\{  P_{\Omega},\mathcal{H}\right\}  =-96\sqrt
{3}e^{-2\sqrt{3}\Omega},\nonumber\\
\dot{\beta}  &  =N\left\{  \beta,\mathcal{H}\right\}  =2P_{\beta},\label{22}\\
\dot{P}_{\beta}  &  =N\left\{  P_{\beta},\mathcal{H}\right\}  =0,\nonumber
\end{align}
where we have used the constraint $h\approx0$ and fixed the gauge $N=\xi
^{-1}=24l^{3}=24\exp\left(  -\sqrt{3}\beta-2\sqrt{3}\Omega\right)  $. From now
on, including the sections on quantum cosmology, we shall restrict all of our
considerations to this gauge. Note that this gauge choice does not correspond
to the comoving cosmological time.

As solutions for $\Omega$ and $\beta$\ we find
\begin{align}
\Omega(t)  &  =\frac{\sqrt{3}}{6}\ln\left\{  \frac{48}{P_{\beta_{0}}^{2}}%
\cosh^{2}\left[  2\sqrt{3}P_{\beta_{0}}\left(  t-t_{0}\right)  \right]
\right\}  ,\nonumber\\
\beta(t)  &  =2P_{\beta_{0}}\left(  t-t_{0}\right)  +\beta_{0}. \label{23.7}%
\end{align}

From (\ref{17}) and (\ref{23.7}) we can evaluate
\begin{align}
\Theta(t)  &  =-\frac{4\sqrt{3}}{P_{\beta_{0}}}\left\{  \cosh^{2}\left[
2\sqrt{3}P_{\beta_{0}}\left(  t-t_{0})\right)  \right]  +\sinh\left[
4\sqrt{3}P_{\beta_{0}}\left(  t-t_{0})\right)  \right]  \right\}  \exp\left[
\sqrt{3}\left(  2P_{\beta_{0}}\left(  t-t_{0}\right)  +\beta_{0}\right)
\right]  ,\nonumber\\
& \nonumber\\
\sigma(t)  &  =\frac{2}{P_{\beta_{0}}}\left\{  4\cosh^{2}\left[  2\sqrt
{3}P_{\beta_{0}}\left(  t-t_{0}\right)  \right]  +\sinh\left[  4\sqrt
{3}P_{\beta_{0}}\left(  t-t_{0}\right)  \right]  \right\}  \exp\left[
\sqrt{3}\left(  2P_{\beta_{0}}\left(  t-t_{0}\right)  +\beta_{0}\right)
\right]  .\label{24}\\
& \nonumber\\
l^{3}\left(  t\right)   &  =\frac{P_{\beta_{0}}^{2}}{48}\operatorname{sech}%
^{2}\left[  2\sqrt{3}P_{\beta_{0}}\left(  t-t_{0}\right)  \right]  \exp\left[
-\sqrt{3}\left(  2P_{\beta_{0}}\left(  t-t_{0}\right)  -\beta_{0}\right)
\right]  .\nonumber
\end{align}

From (\ref{24}), it can be seen that the universe volume expansion $\Theta(t)$
decreases monotonically\ passing by zero at $t=t_{0}-\sqrt{3}\ln\left(
3\right)  /(12P_{\beta_{0}})$. The\ characteristic volume $l^{3}(t)$\ departs
from zero at $t=-\infty,$ increases up to $3\sqrt{3}P_{\beta_{0}}^{2}%
\exp\left[  -\sqrt{3}\beta_{0}\right]  /4$ at $t=t_{0}-\sqrt{3}\ln\left(
3\right)  /12P_{\beta_{0}},$ where it achieves its maximum,\ and then
decreases to zero\footnote{It can be shown that the vanishing of $l^{3}(t)$ at
$t=\pm\infty$ correspond to singularities achieved at finite values of the
cosmic time} at $t=\infty$. The universe shape, on the other hand, departs
from a highly asymmetric state, achieves a configuration of minimum
anisotropy, and again return to a highly asymmetric condition. The typical
behavior of $\Theta(t),$ $\sigma(t)$\ and $l^{3}\left(  t\right)  $\ is
depicted in the thick curves of Figs. $1a,b,$ and $c$ for given values of
$P_{\beta_{0}}$.

\bigskip

\bigskip%

\begin{center}
\includegraphics[
height=3.8839in,
width=6.0234in
]%
{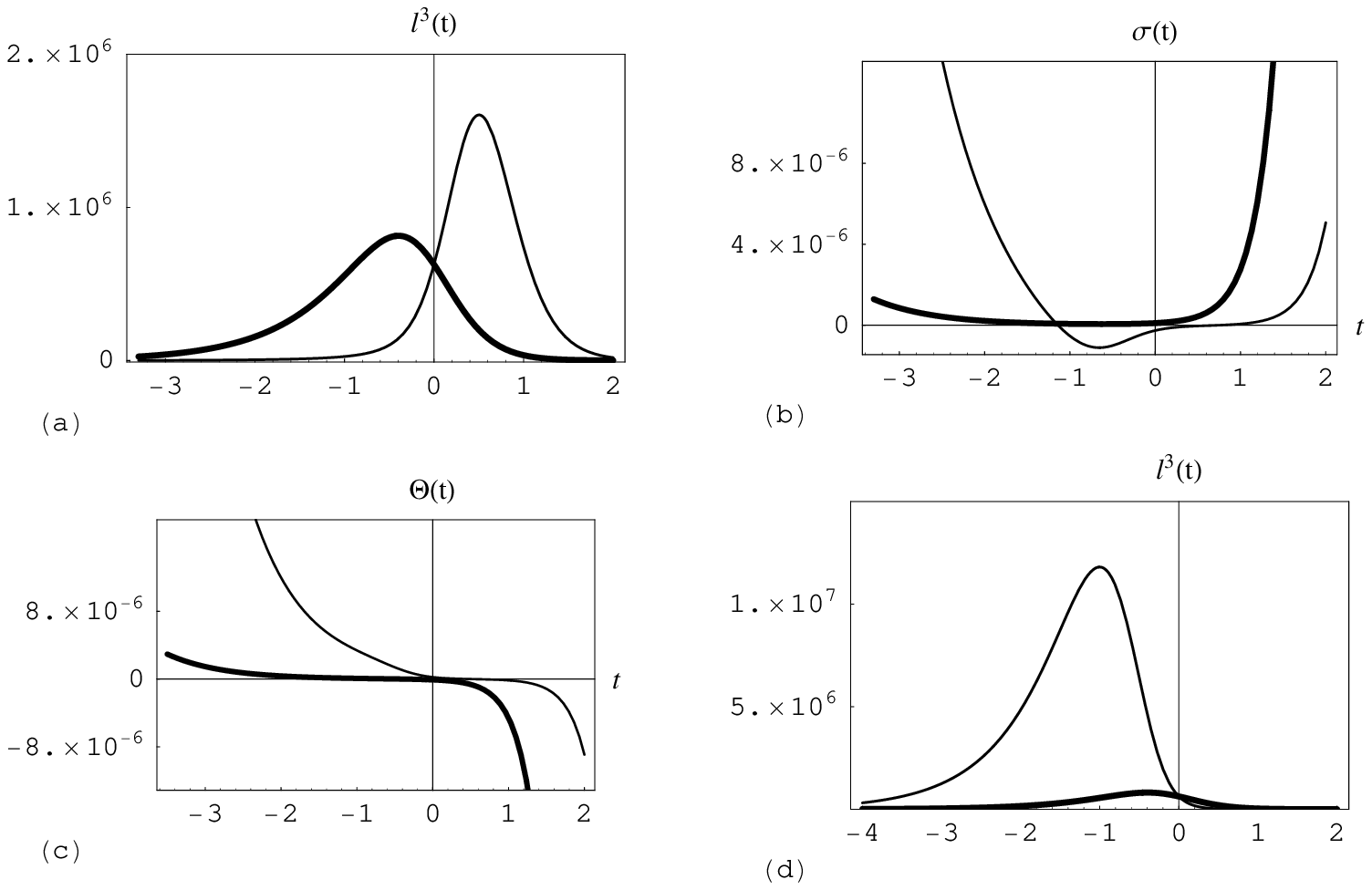}%
\\
FIG.1. The typical behavior of the universe characteristic volume $l^{3}(t)$,
volume expansion $\Theta(t)$, and shear $\sigma(t)$ in the commutative and
noncommutative versions of the classical Kantowski-Sachs universe. The initial
conditions chosen are $\beta_{0}=-11,P_{\beta_{0}}=2/5,t_{0}=0.$ In the plots
$(a)$, $(b)$ and $(c):$ $\theta=0$ (thick lines) and $\theta=5$ (thin lines).
In plot $\left(  d\right)  :$ $\theta=0$ (thick line) and $\theta=-5$ (thin
line).
\end{center}

\section{Noncommutative Classical Model}

The natural step to follow before the introduction of the quantum effects in
the\ Kantowski-Sachs universe is the investigation of the implications of
noncommutative geometry for the classical version of this model.\ Let us
introduce a noncommutative classical geometry in the model by considering a
Hamiltonian that has the same functional form as (\ref{16}) but is valued on
variables that satisfy the deformed Poisson brackets%
\begin{equation}
\left\{  \Omega,P_{\Omega}\right\}  =1\text{,\ \ }\left\{  \beta,P_{\beta
}\right\}  =1\text{,\ \ }\left\{  P_{\Omega},P_{\beta}\right\}  =0\text{,\ \ }%
\left\{  \Omega,\beta\right\}  =\theta. \label{40}%
\end{equation}
The equations of motion in this case are written as
\begin{align}
\dot{\Omega}  &  =-2P_{\Omega},\nonumber\\
\dot{P}_{\Omega}  &  =-96\sqrt{3}e^{-2\sqrt{3}\Omega},\nonumber\\
\dot{\beta}  &  =2P_{\beta}-96\sqrt{3}\theta e^{-2\sqrt{3}\Omega},\label{41}\\
\dot{P}_{\beta}  &  =0.\nonumber
\end{align}
The solutions for $\Omega(t)$ and $\beta(t)$ are
\begin{align}
\Omega(t)  &  =\frac{\sqrt{3}}{6}\ln\left\{  \frac{48}{P_{\beta_{0}}^{2}}%
\cosh^{2}\left[  2\sqrt{3}P_{\beta_{0}}\left(  t-t_{0}\right)  \right]
\right\}  ,\nonumber\\
\beta(t)  &  =2P_{\beta_{0}}\left(  t-t_{0}\right)  +\beta_{0}-\theta
P_{\beta_{0}}\tanh\left[  2\sqrt{3}P_{\beta_{0}}\left(  t-t_{0}\right)
\right]  . \label{42}%
\end{align}
This is the most direct way to obtain the metric variables. A parallel with
the calculation procedure adopted in the quantum case, however, is rendered
easier by the use of the auxiliary canonical variable formalism. We shall
therefore present this alternative approach below.

Instead of working directly with the physical variables $\Omega$ and $\beta$,
we may achieve the solutions above by making use of the auxiliary canonical
variables $\Omega_{c}$ and $\beta_{c},$\ defined as%

\begin{equation}
\Omega_{c}=\Omega+\frac{\theta}{2}P_{\beta}\text{, \ }\beta_{c}=\text{$\beta
-$}\frac{\theta}{2}P_{\Omega}\text{,\ \ }P_{\Omega_{c}}=P_{\Omega}%
\text{,\ \ }P_{\beta_{c}}=P_{\beta}. \label{43}%
\end{equation}
The Poisson brackets for these variables are
\begin{equation}
\left\{  \Omega_{c},P_{\Omega_{c}}\right\}  =1\text{,\ \ }\left\{  \beta
_{c},P_{\beta_{c}}\right\}  =1\text{,\ \ }\left\{  P_{\Omega_{c}},P_{\beta
_{c}}\right\}  =0\text{,\ \ }\left\{  \Omega_{c},\beta_{c}\right\}  =0.
\label{44}%
\end{equation}
As the equations of motion in the canonical formalism in the gauge
$N=24\exp\left(  -\sqrt{3}\beta-2\sqrt{3}\Omega\right)  $ we have
\begin{align}
\dot{\Omega}_{c}  &  =-2P_{\Omega_{c}},\nonumber\\
\dot{P}_{\Omega_{c}}  &  =-96\sqrt{3}e^{-2\sqrt{3}\Omega},\nonumber\\
\dot{\beta}_{c}  &  =2P_{\beta_{c}}-48\sqrt{3}\theta e^{-2\sqrt{3}\Omega
},\label{46}\\
\dot{P}_{\beta_{c}}  &  =0,\nonumber
\end{align}
whose solutions are
\begin{align}
\Omega_{c}(t)  &  =\frac{\sqrt{3}}{6}\ln\left\{  \frac{48}{P_{\beta_{0}}^{2}%
}\cosh^{2}\left[  2\sqrt{3}P_{\beta_{0}}\left(  t-t_{0}\right)  \right]
\right\}  +\frac{\theta}{2}P_{\beta_{0}},\nonumber\\
\beta_{c}(t)  &  =2P_{\beta_{0}}\left(  t-t_{0}\right)  +\beta_{0}%
-\frac{\theta}{2}P_{\beta_{0}}\tanh\left[  2\sqrt{3}P_{\beta_{0}}\left(
t-t_{0}\right)  \right]  ,\nonumber\\
P_{\Omega_{c}}(t)  &  =-P_{\beta_{0}}\tanh\left[  2\sqrt{3}P_{\beta_{0}%
}\left(  t-t_{0}\right)  \right]  ,\label{47}\\
P_{\beta_{c}}(t)  &  =P_{\beta_{0}}.\nonumber
\end{align}
Finally, from (\ref{43}) and (\ref{47}) we can recover the solution (\ref{42}).

Having the noncommutative $\Omega(t)$ and $\beta(t)$ from (\ref{42}), we can
evaluate the expression for $\Theta(t),\sigma(t)$ and $l^{3}\left(  t\right)
$ as%

\begin{align}
\Theta(t)  &  =\left\{  -\frac{4\sqrt{3}}{P_{\beta_{0}}}\left(  \cosh
^{2}\left[  2\sqrt{3}P_{\beta_{0}}\left(  t-t_{0}\right)  \right]
+\sinh\left[  4\sqrt{3}P_{\beta_{0}}\left(  t-t_{0}\right)  \right]  \right)
+4\sqrt{3}\theta\right\} \nonumber\\
&  \times\exp\left[  \sqrt{3}\left(  2P_{\beta_{0}}t+\beta_{0}\right)  -\theta
P_{\beta_{0}}\tanh\left(  2\sqrt{3}P_{\beta_{0}}\left(  t-t_{0}\right)
\right)  \right]  ,\nonumber\\
& \nonumber\\
\sigma(t)  &  =\left\{  \frac{2}{P_{\beta_{0}}}\left(  4\cosh^{2}\left[
2\sqrt{3}P_{\beta_{0}}\left(  t-t_{0}\right)  \right]  +\sinh\left[  4\sqrt
{3}P_{\beta_{0}}\left(  t-t_{0}\right)  \right]  \right)  -4\sqrt{3}%
\theta\right\} \label{49}\\
&  \times\exp\left[  \sqrt{3}\left(  2P_{\beta_{0}}t+\beta_{0}\right)  -\theta
P_{\beta_{0}}\tanh\left(  2\sqrt{3}P_{\beta_{0}}\left(  t-t_{0}\right)
\right)  \right]  ,\nonumber\\
& \nonumber\\
l^{3}\left(  t\right)   &  =\frac{P_{\beta_{0}}^{2}}{48}\operatorname{sech}%
^{2}\left[  2\sqrt{3}P_{\beta_{0}}\left(  t-t_{0}\right)  \right]  \exp\left[
\sqrt{3}\left(  2P_{\beta_{0}}t+\beta_{0}\right)  -\theta P_{\beta_{0}}%
\tanh\left(  2\sqrt{3}P_{\beta_{0}}\left(  t-t_{0}\right)  \right)  \right]
.\nonumber
\end{align}

We start our comparison with the previous case by studying the behavior of the
characteristic volume $l^{3}\left(  t\right)  $. As it can be seen from \ the
expression for $l^{3}\left(  t\right)  $ in\ (\ref{49}), the singular behavior
of the classical noncommutative\ Kantowski-Sachs universe at $t=\pm\infty$ is
the same as that of the classical commutative analog. The $\theta$
contribution in the expression for $\beta(t)$ in (\ref{42})\ does not modify
the behavior of $l^{3}\left(  t\right)  $ in the infinite past or future.
Instead, its contribution is relevant near $t=t_{0}$, where the hyperbolic
tangent varies fast. Near this time, the behavior of $l^{3}\left(  t\right)
$, $\Theta(t)$\ and $\sigma(t)$\ can differ appreciably\ from the commutative
case, as it is shown in the thin curves of Figs. $1a,b,$ and $c$. While the
difference in the behavior of $l^{3}\left(  t\right)  $ and $\Theta(t)$ is
only quantitative, in shear function it is qualitative. The noncommutative
universe can change its expansion directions, and become two times\ isotropic
before being similar in shape to its commutative counterpart at late times. By
varying the initial conditions $\beta_{0}$ and$P_{\beta_{0}},$ the deviation
from the commutative behavior can become very large for $l^{3}\left(
t\right)  $, $\Theta(t)$\ and $\sigma(t)$. For $\beta_{0}=0,$ $P_{\beta_{0}%
}=3$ and $\theta=5$, e.g., we have $l_{\max}^{3}\simeq1.55\cdot10^{8}$, while
for the same initial conditions in the commutative analog we have $l_{\max
}^{3}\simeq0.24.$ Other interesting aspect of the noncommutative universe
solution is its dependence on the sign of the $\theta$ parameter. Fig. $1d$
presents a plot of $l^{3}\left(  t\right)  $ with the same initial conditions
previously adopted in Fig. $1a$, but with the $\theta$ sign inverted. As it
can be seen from the figure, the $l^{3}\left(  t\right)  $ width and maximum
value are considerably enlarged with respect to the similar curve of Fig.
$1a$. Therefore, depending on the initial conditions and on the $\theta$ sign,
it is possible to obtain a appreciable deviation from the commutative behavior
even with smaller values of $\theta$.

\section{Commutative Quantum Model}

Now let us consider the quantum Kantowski-Sachs model. Due to the technical
difficulty in dealing with (\ref{9}), the quantum cosmology is usually based
on the minisuperspace construction of homogeneous models \cite{23,23.5}. With
this approach, it is possible to access a nonperturbative sector of quantum
gravity by paying the price of freezing out infinite degrees of freedom. For
that, an \textit{ansatz} of the type of (\ref{15}) is introduced in (\ref{9}),
and\ the spatial dependence of the metric is integrated out. The
Wheeler-DeWitt equation is thereby reduced to a Klein-Gordon equation. For the
Kantowski-Sachs universe, such an equation is\footnote{We are assuming a
particular factor ordering.}%

\begin{equation}
\left[  -\widehat{P}_{\Omega}^{2}+\widehat{P}_{\beta}^{2}-48\exp\left(
-2\sqrt{3}\Omega\right)  \right]  \Psi(\Omega,\beta)=0, \label{25}%
\end{equation}
where $\widehat{P}_{\Omega}=-i\partial/\partial\Omega$ and $\widehat{P}%
_{\beta}=-i\partial/\partial\beta$. A solution to equation (\ref{25}) is
\cite{15}
\begin{equation}
\Psi_{\nu}(\Omega,\beta)=e^{i\nu\sqrt{3}\beta}K_{i\nu}\left(  4e^{-\sqrt
{3}\Omega}\right)  , \label{26}%
\end{equation}
where $K_{i\nu}$ is a modified Bessel function and $\nu$ is a real\ constant.
Once a quantum state of the universe is given, as, e.g., a superposition of
states\footnote{Since the index in $\nu$ is continuous, in the most general
case\ the sum can be replaced by an integral.}
\begin{equation}
\Psi(\Omega,\beta)=\sum_{\nu}C_{\nu}e^{i\nu\sqrt{3}\beta}K_{i\nu}\left(
4e^{-\sqrt{3}\Omega}\right)  =R\text{ }e^{iS}, \label{27}%
\end{equation}
the universe evolution can be determined by integrating the guiding equation
(\ref{10}). In the minisuperspace approach, the analog of that equation is
\begin{align}
P_{\Omega}  &  =-\frac{1}{2}\dot{\Omega}=\operatorname{Re}\left\{
\frac{\left[  \Psi^{\ast}\left(  -i\hbar\partial_{\Omega}\right)  \Psi\right]
}{\Psi^{\ast}\Psi}\right\}  =\frac{\partial S}{\partial\Omega},\nonumber\\
P_{\beta}  &  =\frac{1}{2}\dot{\beta}=\operatorname{Re}\left\{  \frac{\left[
\Psi^{\ast}\left(  -i\hbar\partial_{\beta}\right)  \Psi\right]  }{\Psi^{\ast
}\Psi}\right\}  =\frac{\partial S}{\partial\beta}. \label{28}%
\end{align}

As before, we have fixed the gauge $N=24l^{3}=24\exp\left(  -\sqrt{3}%
\beta-2\sqrt{3}\Omega\right)  $. Usually, different choices of time yield
different quantum theories \cite{26}. However, when one uses the Bohmian
interpretation in minisuperspace models the situation is identical to that of
the classical case:\footnote{This is not the case beyond minisuperspace. For
details see \cite{23.5}.} different choices yield the same theory \cite{20.5}.
Hence, as long as $l^{3}(t)$ does not pass through zero (which would mean that
the universe has reached a singularity) the above choice for $N(t)$ is valid
for the description of all the universe history.

The minisuperspace analog of the Hamilton Jacobi equation (\ref{12}) is
\begin{equation}
-\frac{1}{24}\left(  \frac{\partial S}{\partial\Omega}\right)  ^{2}+\frac
{1}{24}\left(  \frac{\partial S}{\partial\beta}\right)  ^{2}-2e^{-2\sqrt
{3}\Omega}+\frac{1}{24R}\left(  \frac{\partial^{2}R}{\partial\Omega^{2}}%
-\frac{\partial^{2}R}{\partial\beta^{2}}\right)  =0. \label{29}%
\end{equation}
In what follows, we shall consider some wavefunctions and apply the Bohmian
formalism to investigate the properties of the universe they represent by
means of quantum trajectories.

\subsection{Case 1}

The wavefunction is of the type (\ref{26}). Since the Bessel function
$K_{i\nu}(x)$ is real for $\nu$ real and $x>0$ \cite{26.5}, the phase can be
read directly from the exponential in (\ref{26}): $S=\nu\sqrt{3}\beta$. The
equations of motion are therefore
\begin{equation}
\dot{\Omega}=0,\text{ \ \ \ \ }\dot{\beta}=2\sqrt{3}\nu, \label{30}%
\end{equation}
whose solutions are
\begin{equation}
\Omega=\Omega_{0},\text{ \ \ \ }\beta=2\sqrt{3}\nu\left(  t-t_{0}\right)
+\beta_{0}. \label{31}%
\end{equation}
By substituting (\ref{31}) into (\ref{17}), we can calculate the physical
quantities $\Theta(t),$ $\sigma(t),$ and $l^{3}(t)$ as%

\begin{align}
\Theta(t)  &  =-\frac{\nu}{4}\exp\left[  6\nu\left(  t-t_{0}\right)
+2\sqrt{3}\Omega_{0}+\sqrt{3}\beta_{0}\right]  ,\nonumber\\
\sigma(t)  &  =\frac{\sqrt{3}\nu}{6}\exp\left[  6\nu\left(  t-t_{0}\right)
+2\sqrt{3}\Omega_{0}+\sqrt{3}\beta_{0}\right]  ,\label{32}\\
l^{3}(t)  &  =\exp\left[  -6\nu\left(  t-t_{0}\right)  -2\sqrt{3}\Omega
_{0}-\sqrt{3}\beta_{0}\right]  .\nonumber
\end{align}

From (\ref{32}) it is easy to see that, according to the sign of $\nu$, there
are two possibilities for the universe evolution. The first $\left(
\nu>0\right)  $ corresponds to a universe that starts with infinitely large
and isotropic volume in the remote past and evolves contracting to a
configuration of small and distorted volume. The second $\left(  \nu>0\right)
$\ is a universe whose volume is infinitely small and distorted in the remote
past,\ and evolves expanding to a large and isotropic configuration in the
infinite future. This qualitative different behavior from the classical
counterpart can be intuitively understood by evaluating the quantum potential.
From (\ref{14}) and (\ref{26}) we can calculate
\begin{equation}
Q=\frac{1}{24R}\left(  \frac{\partial^{2}R}{\partial\Omega^{2}}-\frac
{\partial^{2}R}{\partial\beta^{2}}\right)  =2e^{-2\sqrt{3}\Omega}-\frac
{\nu^{2}}{8}. \label{33}%
\end{equation}
Since $Q$ does not depend on $\beta$, we expect this variable to have a
classical behavior, while $\Omega$\ should encode all the quantum effects.
This is the main reason for the solution for $\beta$ in\ (\ref{31}) be exactly
equal to its classical counterpart (\ref{23.7}) if we identify $P_{\beta_{0}%
}=\sqrt{3}\nu$, while $\Omega=\Omega_{0}$ is radically different.

\subsection{Case 2}

The wavefunction is a superposition of two solutions of the type (\ref{26}),%

\begin{equation}
\Psi(\Omega,\beta)=A_{1}K_{i\mu}\left(  4e^{-\sqrt{3}\Omega}\right)
e^{i\sqrt{3}\mu\beta}+A_{2}K_{i\nu}\left(  4e^{-\sqrt{3}\Omega}\right)
e^{i\sqrt{3}\nu\beta}. \label{34}%
\end{equation}
The corresponding phase is%

\begin{equation}
S(\Omega,\beta)=\arctan\left[  \frac{A_{1}K_{i\mu}\left(  4e^{-\sqrt{3}\Omega
}\right)  \sin\left(  \sqrt{3}\mu\beta\right)  +A_{2}K_{i\nu}\left(
4e^{-\sqrt{3}\Omega}\right)  \sin\left(  \sqrt{3}\nu\beta\right)  }%
{A_{1}K_{i\mu}\left(  4e^{-\sqrt{3}\Omega}\right)  \cos\left(  \sqrt{3}%
\mu\beta\right)  +A_{2}K_{i\nu}\left(  4e^{-\sqrt{3}\Omega}\right)
\cos\left(  \sqrt{3}\nu\beta\right)  }\right]  , \label{35}%
\end{equation}
where the $A_{1}$ and $A_{2}$\ are chosen\ as real coefficients. The equations
of motion (\ref{28}) for this state are
\begin{align}
\frac{d\Omega}{dt}  &  =8\sqrt{3}\text{ }\frac{A_{1}A_{2}\left[  K_{i\mu
}^{\prime}K_{i\nu}-K_{i\mu}K_{i\nu}^{\prime}\right]  \exp\left[  -\sqrt
{3}\Omega\right]  \sin\left[  \sqrt{3}\left(  \mu-\nu\right)  \beta\right]
}{\left(  A_{1}K_{i\mu}\right)  ^{2}+\left(  A_{2}K_{i\nu}\right)  ^{2}%
+2A_{1}A_{2}K_{i\mu}K_{i\nu}\cos\left[  \sqrt{3}\left(  \mu-\nu\right)
\beta\right]  },\nonumber\\
& \label{36}\\
\frac{d\beta}{dt}  &  =2\sqrt{3}\text{ }\frac{\mu A_{1}^{2}K_{i\mu}^{2}+\nu
A_{2}^{2}K_{i\nu}^{2}+\left(  \mu+\nu\right)  A_{1}A_{2}K_{i\mu}K_{i\nu}%
\cos\left[  \sqrt{3}\left(  \mu-\nu\right)  \beta\right]  }{\left(
A_{1}K_{i\mu}\right)  ^{2}+\left(  A_{2}K_{i\nu}\right)  ^{2}+2A_{1}%
A_{2}K_{i\mu}K_{i\nu}\cos\left[  \sqrt{3}\left(  \mu-\nu\right)  \beta\right]
},\nonumber
\end{align}
where prime means derivative with respect to the argument.

The system (\ref{36}) constitutes an autonomous set of nonlinear coupled
differential equations. Although it is hard to solve analytically this system,
the global properties\ of the solutions\ can be easily grasped by considering
the associated field of velocities. A first inspection on the RHS
of\ (\ref{36}) reveals that the velocity field has its direction inverted by
the\ replacement $\mu\rightarrow-\mu$ , $\nu\rightarrow-\nu$. Therefore, to
have a qualitative picture of the velocity field, it is sufficient to consider
$\mu>0$ and study the cases where $\nu>0$ and $\nu<0.$ For simplicity, let us
fix $A_{1}=A_{2}=1/\sqrt{2}$. The most interesting of the two cases, $\nu<0,$
gives rise to the velocity field that is depicted normalized\ in Fig. $2$.

\bigskip%

\begin{center}
\includegraphics[
height=3.5924in,
width=4.4408in
]%
{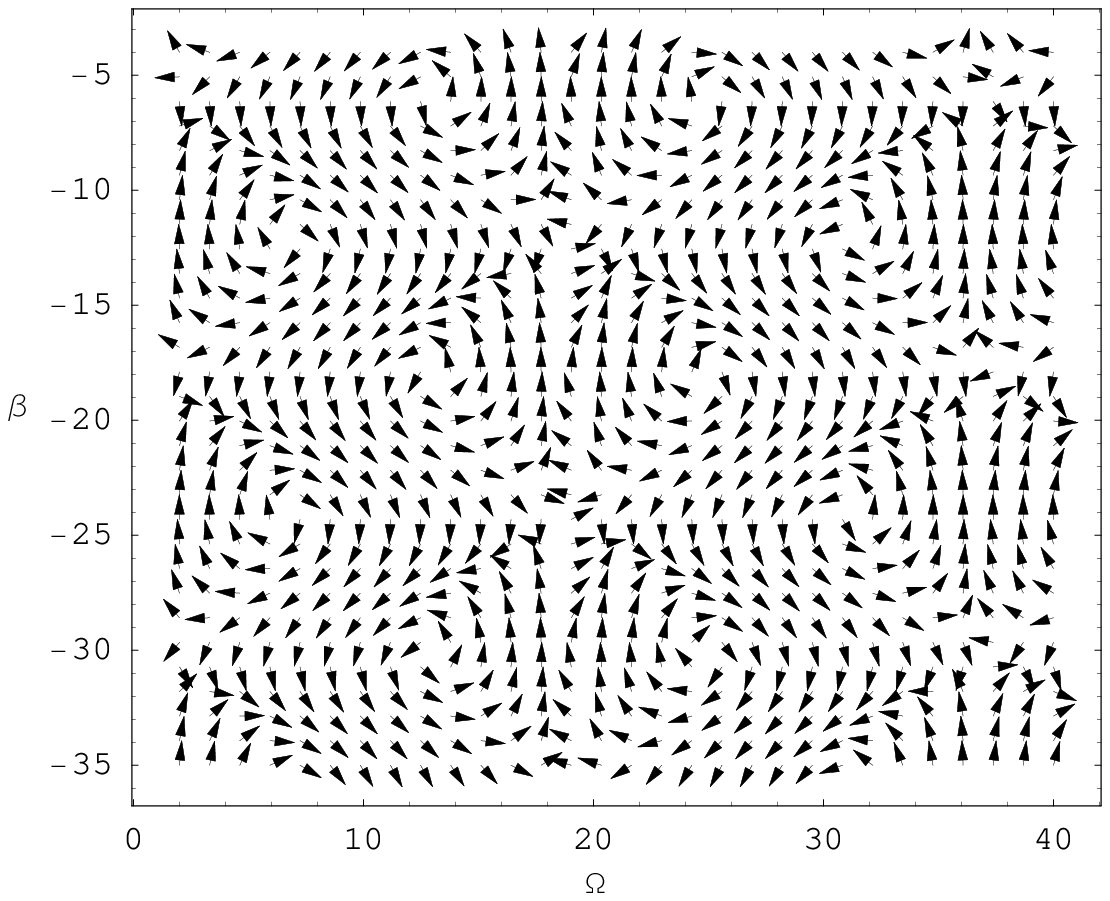}%
\\
FIG. 2. The normalized field of velocities corresponding to the Bohmian
differential equations for the commutative Kantowski-Sachs universe with
$\mu=1/10$ and $\nu=-1/5$.
\end{center}

\bigskip

A simple inspection on (\ref{36}) suggests that there should exist stability
points appearing periodically with period $2\pi/\sqrt{3}\left(  \mu
-\nu\right)  $ along the$\ \beta$ direction. Indeed, for the values given to
$\mu$ and $\nu$ in the plots of Fig. $2$ we have $2\pi/\sqrt{3}\left\vert
\mu-\nu\right\vert =12.09$, which matches exactly with the observed period of
appearance of the stability points along $\beta$ direction.

By varying the initial conditions and the values of $\mu$ and $\nu,$ we can
find a great variety of solutions of (\ref{36}). In what follows we shall
exhibit some of them, giving preference for the ones which correspond to
non-singular universes.

From the field plot of Fig. $2$ we\ can obtain information about the
qualitative behavior of the solutions for $l^{3}(t)$, $\Theta(t)$ and
$\sigma(t)$. If $\Omega(t)$ and $\beta(t)$ are periodic, equations (\ref{17})
show that $l^{3}(t)$, $\Theta(t)$ and $\sigma(t)$\ have the same behavior.
This is the solution type we expect to find when integrating the system
(\ref{36}) around the stability point near $\Omega=6$ and$\ \beta=-24$, for
example. In Fig. $2$ it can also be seen, from the flow emerging around
$\Omega=11$ or $\Omega=27$, that there exist solutions monotonically
increasing in $\beta$ and oscillatory in $\Omega.$ In addition to the periodic
universes, we therefore expect to find universe solutions that contract
(expand) with $l^{3}(t)$ passing by a sequence of bounces and becoming
singular in the infinite future (past).

We start our numerical study with the periodic solutions. In order to find
them, we shall focus our attention on the vertical straight line that crosses
the $\Omega$ axes\ near $\Omega=6$ in Fig. $2$. Without loss of generality,
let us consider the\ stability point located near $\Omega=6$ and $\beta=-24$.
By solving (\ref{36}) numerically with initial conditions $\Omega(0)=2$ and
$\beta(0)=-24$ and computing the respective $\ln\left[  l^{3}(t)\right]  $,
$\Theta(t)$ and $\sigma(t),$ we verify\ that these three quantities indeed
present a periodic behavior (Figs. $3a,c,$ and $e$). An interesting
information that can be read directly from the plot of $\ln\left[
l^{3}(t)\right]  $ in Fig. $3a$\ is the number of $e$-folds between the
maximum and minimum universe volumes ($\simeq$ $29$ $e$-folds). It is easy to
see that the number of $e$-folds and the value of the minimum volume can be
adjusted by varying the initial conditions or changing the stability point.
For a larger number of $e$-folds, it is enough to enlarge the orbit radius by
choosing $\Omega(0)$ and $\beta(0)$ appropriately.

By studying the $l^{3}\left(  t\right)  $ function, we can know how many times
the universe volume is larger then the Planckian volume $l_{p}^{3}$. In the
unit system adopted, $l_{p}^{3}\sim10^{-3}$. From (\ref{17}) we can see that
$l_{\min}^{3}$\ can be rendered\ larger (smaller) if we decrease (increase),
e.g., $\beta_{\max}$. This is accomplished by\ changing to a similar orbit
around a stability point immediately below (above) moving\ vertically along
the $\beta$ direction. If one keeps $\Omega(0)$ unaltered and decreases
$\beta(0)$ by the spacing between$\ $the orbit centers, the difference in
$\beta_{\max}$ will be decreased or increased about $12$. The corresponding
decrease or increase in $l_{\min}^{3}$ will therefore be about $\exp
[12\sqrt{3}]\simeq21$ $e$-folds.

In addition to $l^{3}(t)$, the variables $\Theta(t)$ and $\sigma(t)$ provide
relevant information about the universe behavior\ during each of its periodic
cycles. A general inspection on Fig. $3$ reveals that\ the universe in
question alternates between configurations of large volume, almost uniform
shape and small volume\ expansion, with configurations of small volume,
distorted shape and large volume expansion. In each of its cycles the universe
is isotropic at two times (Figs $3g$ and $h$).

Let us now turn our attention to the flow emerging around $\Omega=11$ or
$\Omega=27$ in Fig. $2$. The solutions in these regions correspond to
universes that start at $t=-\infty$ with infinite volume and contract passing
by a sequence of bounces up to a singularity, as is shown in the logarithmic
plot of $l^{3}(t)$ in\ Fig. $3b$. The logarithms of the\ corresponding volume
expansion and shear appear plotted in Figs. $3d$ and $3f$, respectively. In
the same way as in the periodic solutions, the regions where the universe is
small correspond exactly to the ones of maximum anisotropy. This is verified
by moving from the local minimums at Fig. $3b$ and moving down vertically to
arrive near the local maximums at Fig. $3d$.\ The small creases in the top of
the picks in Fig. $3f$ account for the abrupt change in direction of the
expansion that occur in each of the bouncing regions. As stated before,
the\ velocity field in Fig. $2$, has its direction inverted whenever
the\ replacement $\mu\rightarrow-\mu$, $\nu\rightarrow-\nu$ is made. We can
use this property to construct an expanding solution where $l^{3}(t)$ starts
from a singularity and increases up to infinity\ passing by a sequence of
bounces from the solution depicted in Figs. $3b,d$ and $f$.

Another different solution type is present in the case where $\nu=-\mu$. The
phase of such a state is $S=0$. We have therefore an static universe of
arbitrary size, a genuinely quantum behavior. Indeed the quantum potential for
this state,
\[
Q=2e^{-2\sqrt{3}\Omega},
\]
cancels exactly the classical potential. This justifies the highly
nonclassical behavior observed.

\newpage%

\begin{center}
\includegraphics[
height=7.8473in,
width=6.0226in
]%
{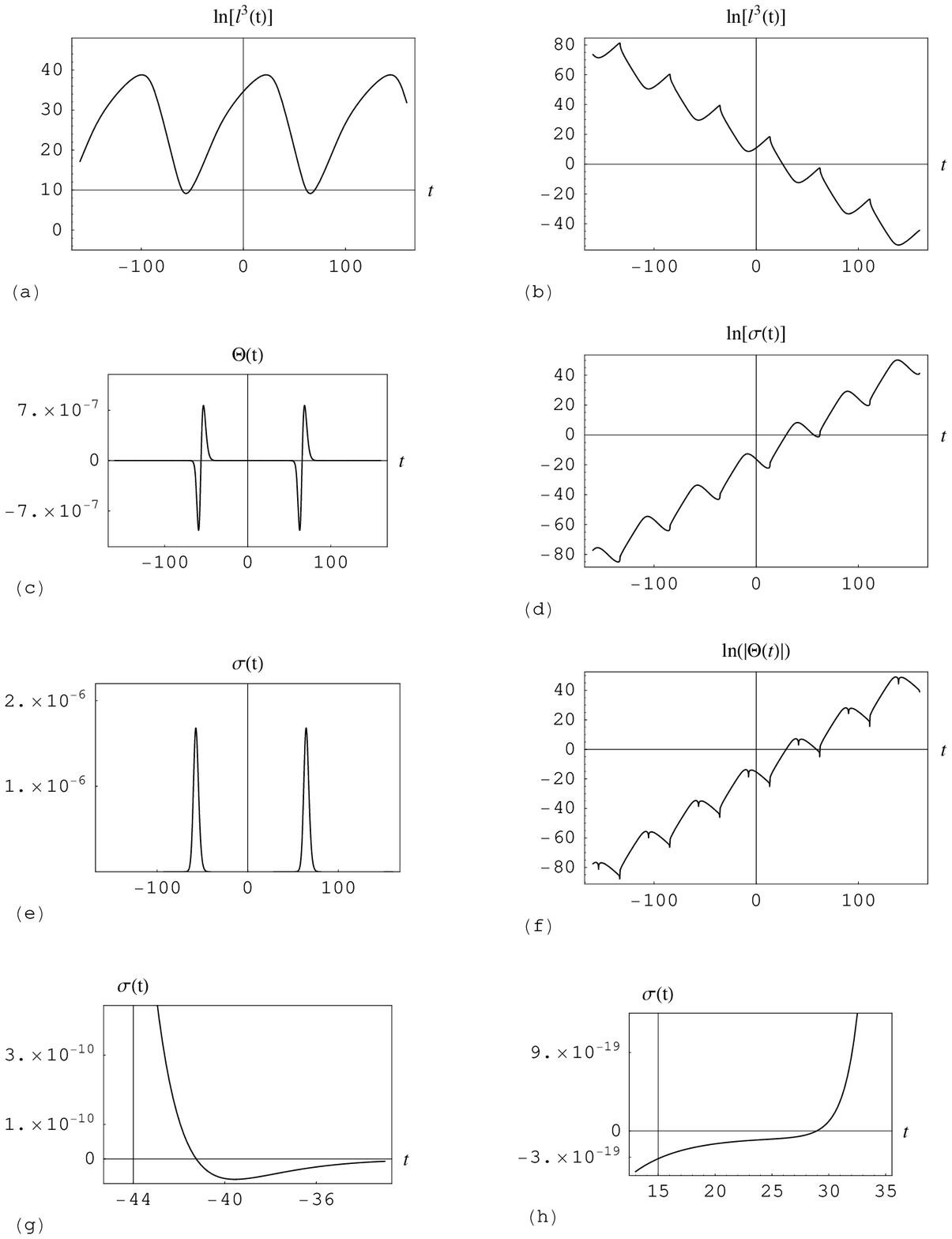}%
\\
FIG. 3. The evolution of the universe characteristic volume, volume expansion,
and shear in the commutative quantum Kantowski-Sachs universe with $\mu=1/10$
and $\nu=-1/5.$ $(a),(c),$ and $(e):$ $\Omega(0)=2,\beta(0)=-24.$ $(b),(d),$
and $(f):$ $\Omega(0)=7.7,\beta(0)=-21.7.$ $\left(  g\right)  $ and $\left(
h\right)  $ : enlarged plots\ of parts of $\left(  e\right)  .$%
\end{center}

\bigskip

All the solutions discussed in this section are interesting in their own by
the mathematically allowed universes they represent. From the phenomenological
point of view, however, it is interesting to know that there exist
non-singular\ (periodic) dynamic solutions that can account for our
expectation that the minimum length achieved by the universe be larger than
the Planckian one. Near this length scale, the effective quantum gravity
theory based on the Wheeler -DeWitt equation is no more expected to be valid.
For the solution plotted in Fig. $3a$, e.g., $l_{\min}\simeq20\sim200l_{P}$.

\section{The Noncommutative Quantum Model}

Having studied the quantum and noncommutativity effects in isolated examples,
we now have joined elements to analyze the combination of them, which is
realized in the noncommutative quantum model. In the construction\ of this
model, we shall follow a prescription similar to that proposed in \cite{15},
where a canonical deformation in the algebra of the minisuperspace operators
is introduced. As in the ordinary quantum mechanics, such a kind of
deformation is usually defined with respect to a \textquotedblleft
preferred\textquotedblright\ set of Cartesian coordinates, where the
noncommutative parameter is taken to be constant.\ As this preferred set, we
shall take the one constituted by the configuration variables $\Omega$ and
$\beta$,%

\begin{equation}
\lbrack\widehat{\Omega},\widehat{\beta}]=i\theta. \label{50}%
\end{equation}

According to the Weyl quantization procedure \cite{2,3}, the realization of
the commutation relation (\ref{50}) between the observables $\widehat{\Omega}
$ and $\widehat{\beta}$ in terms of commutative functions is made by
the\ Moyal star product, defined as below%

\begin{equation}
f(\Omega_{c},\beta_{c})\star g(\Omega_{c},\beta_{c})=f(\Omega_{c},\beta
_{c})e^{i\frac{\theta}{2}\left(  \overleftarrow{\partial}_{\Omega_{c}%
}\overrightarrow{\partial}_{\beta_{c}}-\text{ }\overleftarrow{\partial}%
_{\beta_{c}}\overrightarrow{\partial}_{\Omega_{c}}\right)  }g(\Omega_{c}%
,\beta_{c}). \label{51}%
\end{equation}
The commutative coordinates $\Omega_{c}$ and $\beta_{c}$ are called Weyl
symbols of the operators $\widehat{\Omega}$ and $\widehat{\beta}$,
respectively. The notation here is suggestive because these symbols match
exactly with the canonical variables as defined by (\ref{43}), which may be
seen directly from the properties of the Moyal product.

In order to compare evolutions with the same time parameter as in the previous
cases, we again fix the gauge $N=24\exp\left(  -\sqrt{3}\beta-2\sqrt{3}%
\Omega\right)  $ in (\ref{16}). The Wheeler-DeWitt equation for the
noncommutative Kantowsky-Sachs model is \cite{15}
%

\begin{equation}
\left[  -P_{\Omega_{c}}^{2}+P_{\beta_{c}}^{2}-48\exp\left(  -2\sqrt{3}%
\Omega_{c}\right)  \right]  \star\Psi(\Omega_{c},\beta_{c})=0, \label{52}%
\end{equation}
which is the Moyal deformed version of (\ref{25}). By using the properties of
the Moyal product, it is possible to write the potential term (which we denote
by $V$ to include the general case) as
\begin{align}
V(\Omega_{c},\beta_{c})\star\Psi(\Omega_{c},\beta_{c})  &  =V\left(
\Omega_{c}+i\frac{\theta}{2}\partial_{\beta_{c}},\beta_{c}-i\frac{\theta}%
{2}\partial_{\Omega_{c}}\right)  \Psi(\Omega_{c},\beta_{c})\nonumber\\
&  =V\left(  \widehat{\Omega},\widehat{\beta}\right)  \Psi(\Omega_{c}%
,\beta_{c}) \label{53}%
\end{align}
where%

\begin{equation}
\widehat{\Omega}=\widehat{\Omega}_{c}-\frac{\theta}{2}\widehat{P}_{\beta_{c}%
}\text{, \ \ }\widehat{\beta}=\widehat{\text{$\beta$}}\text{$_{c}$}%
+\frac{\theta}{2}\widehat{P}_{\Omega_{c}}. \label{54}%
\end{equation}
Equation (\ref{54}) is nothing but the operatorial version of
equation\ (\ref{43}). The Wheeler-DeWitt equation then reads
\begin{equation}
\left[  -\widehat{P}_{\Omega_{c}}^{2}+\widehat{P}_{\beta_{c}}^{2}%
-48\exp\left(  -2\sqrt{3}\widehat{\Omega}_{c}+\sqrt{3}\theta\widehat{P}%
_{\beta_{c}}\right)  \right]  \Psi(\Omega_{c},\beta_{c})=0. \label{55}%
\end{equation}

Two consistent interpretations for the cosmology which emerges from equations
(\ref{50})-(\ref{53}) are possible. The first consists in considering
the\ Weyl symbols $\Omega_{c}$ and $\beta_{c}$ as the constituents of the
physical metric. In this case the theory is essentially commutative with a
modified interaction. In\ the second interpretation, which is adopted, e.g.,
in \cite{8,9}, the Weyl symbols are considered as auxiliary coordinates, in
the same way as in the classical case discussed in the previous section. Such
an interpretation is closer to the spirit of this work, which is to study the
evolution of a noncommutative quantum universe. Since it is the algebra of
$\widehat{\Omega}$ and $\widehat{\beta}$, rather than the algebra of
$\widehat{\Omega}_{c}$ and $\widehat{\beta}_{c},$ that satisfies (\ref{50}),
we shall interpret the $\widehat{\Omega}$ and $\widehat{\beta}$ as the
operators associated with the physical metric. Moreover, the adoption of
$\widehat{\Omega}$ and $\widehat{\beta}$ as the operators associated with the
physical metric\ is also in accordance with the Dirac quantization procedure
\[
\left\{  \text{ },\text{ }\right\}  \rightarrow\frac{1}{i}\left[  \text{
},\text{ }\right]  ,
\]
if one departs from the noncommutative classical analog discussed before.\ In
the context of ordinary quantum mechanics, the two points of view provide the
same energy spectrum, which is the physical quantity calculated in many works
(see, e.g., \cite{39}). However, as long as one wants to give an ontology to
the theory (to circumvent fundamental problems of quantum cosmology), a
precise specification of the objects the theory refers to must be made.

\subsection{Bohmian Formalism for Noncommutative Minisuperspaces}

In order to go on in our comparative study of the Kantowski-Sachs universe, it
is important to develop a Bohmian formulation for noncommutative quantum
cosmology. This will be carried out here in the context of the minisuperspace
formalism, which is our systematic tool in\ the study of quantum cosmological models.

In Bohmian noncommutative quantum cosmology we want to deal with a formalism
that allow us to trace a clear picture of the universe evolution in a similar
way as in the commutative quantum cosmology. One could ask how is this
possible, since we are dealing with noncommutative coordinates that satisfy
(\ref{50}). The answer is that the operatorial formalism of quantum mechanics
with operators acting in a Hilbert space of states\ is not a primary concept
in Bohmian quantum mechanics. This is exactly one of the features of Bohmian
formalism that renders it interesting for application in quantum cosmology,
where there is no external observer. In commutative Bohmian quantum mechanics
it is possible to describe particles with well defined position and momentum
at each instant of time, although their position and momentum operators
satisfy (for details see \cite{20})
\begin{equation}
\lbrack\widehat{x}^{i},\widehat{p}^{j}]=i\hbar\delta^{ij}. \label{55.5}%
\end{equation}
Thus, it is reasonable to expect that in Bohmian noncommutative quantum
cosmology it should be possible to describe the metric variables as well
defined entities, although the operators $\widehat{\Omega}$ and $\widehat
{\beta}$ satisfy (\ref{50}). Indeed, this is exactly the case in the
formulation proposed here.\footnote{For a related work on the Bohmian
interpretation in the context of ordinary non-relativistic quantum mechanics
see \cite{9}.}

The key ingredients in our Bohmian formalism are the wavefunction, which
contain information about the universe evolution, and the metric variables
$\Omega$ and $\beta$. To these quantities we want to give an objective
meaning. The wavefunction can be obtained by solving (\ref{52}). What is
missing therefore is the evolution law for $\Omega$ and $\beta$. A simple and
direct way to find this evolution law is by extending the formalism of section
6 employing the mapping described below.

To the\ Hermitian operator $\widehat{A}(\widehat{\Omega}_{c},\widehat{\beta
}_{c},\widehat{P}_{\Omega_{c}},\widehat{P}_{\beta_{c}})$ it is possible to
associate a function $\mathcal{A}\left(  \Omega_{c},\beta_{c}\right)  $
according to the rule
\begin{equation}
\mathcal{B}[\widehat{A}]=\frac{\operatorname{Re}\left[  \Psi^{\ast}\left(
\Omega_{c},\beta_{c}\right)  \widehat{A}\left(  \Omega_{c},\beta_{c}%
,-i\hbar\partial_{\Omega_{c}},-i\hbar\partial_{\beta_{c}}\right)  \Psi\left(
\Omega_{c},\beta_{c}\right)  \right]  }{\Psi^{\ast}\left(  \Omega_{c}%
,\beta_{c}\right)  \Psi\left(  \Omega_{c},\beta_{c}\right)  }=\mathcal{A}%
\left(  \Omega_{c},\beta_{c}\right)  , \label{56}%
\end{equation}
where the real value was taken to account for the hermiticity of $\widehat{A}%
$. The operation (\ref{56}) could be called \textquotedblleft beable
mapping\textquotedblright\footnote{In the context of ordinary non-relativistic
quantum mechanics, where a probability intepretation can be given to
$\rho=\Psi^{\ast}\Psi$, the same procedure is know as \textquotedblleft taking
the local\ expectation value\textquotedblright\ \cite{20}. Such a nomenclature
is clearly senseless here, where $\rho=\Psi^{\ast}\Psi$ does not have a
probabilistic interpretation.}, since it associates with each Hermitian
operator $\widehat{A}$ its corresponding \textquotedblleft
beable\textquotedblright, the\ element of reality (ontology) that lies behind
$\widehat{A}$ in the Bohmian approach.\footnote{Although the operators
$\widehat{\Omega}$ and $\widehat{\beta}$ do not commute, we shall refer
to$\ \Omega$ and $\beta$ as their respective \textquotedblleft
beables\textquotedblright\cite{16.5} by the ontology of the spacetime metric
encoded in these variables.}

By applying (\ref{56}) to evaluate the beables corresponding to the operators
$\widehat{\Omega}$ and $\widehat{\beta}$ we find\
\begin{equation}
\Omega\left(  \Omega_{c},\beta_{c}\right)  =\mathcal{B}[\widehat{\Omega
}]=\frac{\operatorname{Re}\left[  \Psi^{\ast}\left(  \Omega_{c},\beta
_{c}\right)  \widehat{\Omega}(\Omega_{c},-i\hbar\partial_{\beta_{c}}%
)\Psi\left(  \Omega_{c},\beta_{c}\right)  \right]  }{\Psi^{\ast}\left(
\Omega_{c},\beta_{c}\right)  \Psi\left(  \Omega_{c},\beta_{c}\right)  }%
=\Omega_{c}-\frac{\theta}{2}\partial_{\beta_{c}}S \label{57}%
\end{equation}%
\begin{equation}
\beta\left(  \Omega_{c},\beta_{c}\right)  =\mathcal{B}[\widehat{\beta}%
]=\frac{\operatorname{Re}\left[  \Psi^{\ast}\left(  \Omega_{c},\beta
_{c}\right)  \widehat{\beta}(\beta_{c},-i\hbar\partial_{\Omega_{c}}%
)\Psi\left(  \Omega_{c},\beta_{c}\right)  \right]  }{\Psi^{\ast}\left(
\Omega_{c},\beta_{c}\right)  \Psi\left(  \Omega_{c},\beta_{c}\right)
}=\text{$\beta_{c}$}+\frac{\theta}{2}\partial_{\Omega_{c}}S. \label{58}%
\end{equation}

The strategy to find $\Omega\left(  t\right)  $ and $\beta\left(  t\right)  $
now becomes clear. The relevant information for universe evolution can be
extracted from the guiding wave $\Psi\left(  \Omega_{c},\beta_{c}\right)  $ by
first\ computing the associated canonical position tracks $\Omega_{c}\left(
t\right)  $ and $\beta_{c}\left(  t\right)  $. After that, we can obtain
$\Omega\left(  t\right)  $ and $\beta\left(  t\right)  $ by\ evaluating
(\ref{57}) and (\ref{58}) at $\Omega_{c}=\Omega_{c}\left(  t\right)  $
and$\ \beta_{c}=\beta_{c}\left(  t\right)  $, in a close similarity with the
procedure of the second route to calculate $\Omega\left(  t\right)  $ and
$\beta\left(  t\right)  $ in section 5.

Differential equations for the canonical positions $\Omega_{c}\left(
t\right)  $ and $\beta_{c}\left(  t\right)  $ may be found by identifying
$\dot{\Omega}_{c}\left(  t\right)  $\ and $\dot{\beta}_{c}\left(  t\right)  $
with the\ the beables associated with\ their time evolution. In our time gauge
$N=24\exp\left(  -\sqrt{3}\beta-2\sqrt{3}\Omega\right)  $, the Hamiltonian
$H=N\xi h$, with $\xi$ and $h$ defined in Eq. (\ref{20}), reduces simply to
$h$. We can therefore use $h$ to generate time displacements and obtain the
equations of motion for $\Omega_{c}\left(  t\right)  $ and $\beta_{c}\left(
t\right)  $ as%

\begin{equation}
\dot{\Omega}_{c}\left(  t\right)  =\left.  \mathcal{B}\left(  \frac{1}%
{i}[\widehat{\Omega}_{c},\widehat{h}]\right)  \right\vert _{\substack{\Omega
_{c}=\Omega_{c}(t)\\\beta_{c}=\beta_{c}(t)}}=\left.  -2\frac{\partial
S}{\partial\Omega_{c}}\right\vert _{\substack{\Omega_{c}=\Omega_{c}%
(t)\\\beta_{c}=\beta_{c}(t)}}, \label{61}%
\end{equation}

\begin{equation}
\dot{\beta}_{c}\left(  t\right)  =\left.  \mathcal{B}\left(  \frac{1}%
{i}[\widehat{\beta}_{c},\widehat{h}]\right)  \right\vert _{\substack{\Omega
_{c}=\Omega_{c}(t)\\\beta_{c}=\beta_{c}(t)}}=\left.  \left[  2\frac{\partial
S}{\partial\beta_{c}}-48\sqrt{3}\theta\operatorname{Re}\left\{  \frac
{\exp\left(  -2\sqrt{3}\Omega_{c}-i\theta\sqrt{3}\partial_{\beta_{c}}\right)
\left(  R\text{ }e^{iS}\right)  }{R\text{ }e^{iS}}\right\}  \right]
\right\vert _{\substack{\Omega_{c}=\Omega_{c}(t)\\\beta_{c}=\beta_{c}(t)}}.
\label{62}%
\end{equation}
As long as $\Omega_{c}\left(  t\right)  $ and $\beta_{c}\left(  t\right)  $
are known, the\ minisuperspace trajectories are given by
\begin{equation}
\Omega\left(  t\right)  =\Omega_{c}\left(  t\right)  -\frac{\theta}{2}%
\partial_{\beta_{c}}S\left[  \Omega_{c}\left(  t\right)  ,\beta_{c}\left(
t\right)  \right]  , \label{63}%
\end{equation}%
\begin{equation}
\beta\left(  t\right)  =\text{$\beta_{c}$}\left(  t\right)  +\frac{\theta}%
{2}\partial_{\Omega_{c}}S\left[  \Omega_{c}\left(  t\right)  ,\beta_{c}\left(
t\right)  \right]  . \label{64}%
\end{equation}
A solution to (\ref{52}) is \cite{15}
\begin{equation}
\Psi_{\nu}(\Omega_{c},\beta_{c})=e^{i\nu\sqrt{3}\beta_{c}}K_{i\nu}\left\{
4\exp\left[  -\sqrt{3}\left(  \Omega_{c}-\frac{\sqrt{3}}{2}\nu\theta\right)
\right]  \right\}  . \label{65}%
\end{equation}
Once a quantum state of the universe is given as, e.g., a superposition of
states
\begin{equation}
\Psi(\Omega_{c},\beta_{c})=\sum_{\nu}C_{\nu}e^{i\nu\sqrt{3}\beta_{c}}K_{i\nu
}\left\{  4\exp\left[  -\sqrt{3}\left(  \Omega_{c}-\frac{\sqrt{3}}{2}\nu
\theta\right)  \right]  \right\}  =R\text{ }e^{iS}, \label{66}%
\end{equation}
the universe evolution can be determined by solving the system of equations
constituted by (\ref{61}) and (\ref{62}) and substituting the solution in
(\ref{63}) and (\ref{64}).

Before applying the formalism to practical calculations, it is interesting to
understand the meaning of all the terms in equation (\ref{62}).\ This is
accomplished by considering the associated Hamilton-Jacobi equation. The
generalized Hamilton-Jacobi equation for noncommutative quantum cosmology is
obtained by plugging $\Psi(\Omega,\beta)=R$ $e^{iS}$ into (\ref{52}) and
taking the real piece. As a result, we find
\begin{equation}
-\frac{1}{24}\left(  \frac{\partial S}{\partial\Omega_{c}}\right)  ^{2}%
+\frac{1}{24}\left(  \frac{\partial S}{\partial\beta_{c}}\right)
^{2}+V+V_{nc}+Q_{K}+Q_{I}=0, \label{67}%
\end{equation}
where
\begin{align}
V  &  =-2e^{-2\sqrt{3}\Omega_{c}},\nonumber\\
V_{nc}  &  =2e^{-2\sqrt{3}\Omega_{c}}-2e^{-2\sqrt{3}\Omega_{c}+\sqrt{3}%
\theta\partial_{\beta_{c}}S},\nonumber\\
Q_{K}  &  =\frac{1}{24R}\left(  \frac{\partial^{2}R}{\partial\Omega_{c}^{2}%
}-\frac{\partial^{2}R}{\partial\beta_{c}^{2}}\right)  ,\label{68}\\
Q_{I}  &  =-2\operatorname{Re}\left\{  \frac{\exp\left(  -2\sqrt{3}\Omega
_{c}-i\theta\sqrt{3}\partial_{\beta_{c}}\right)  \left(  R\text{ }%
e^{iS}\right)  }{R\text{ }e^{iS}}\right\}  +2e^{-\left(  2\sqrt{3}\Omega
_{c}+\theta\sqrt{3}\partial_{\beta_{c}}S\right)  }.\nonumber
\end{align}
The term $V_{nc}$ is the noncommutative part of the classical potential,
while$\ Q_{K}$ and\ $Q_{I}$\ are denominated as the kinetic and interaction
quantum potentials \cite{8,9}.\ Due to the noncommutative corrections to the
wavefunction, it is clear that, although functionally similar, $Q_{K}$ differs
from $Q$ previously defined in section 4. Equation (\ref{62}) can now be
written as
\begin{equation}
\dot{\beta}_{c}\left(  t\right)  =\left.  \left[  2\frac{\partial S}%
{\partial\beta_{c}}-48\sqrt{3}\theta e^{-2\sqrt{3}\Omega_{c}}+24\sqrt{3}%
\theta\left(  V_{nc}+Q_{I}\right)  \right]  \right\vert _{\substack{\Omega
_{c}=\Omega_{c}(t)\\\beta_{c}=\beta_{c}(t)}}. \label{69}%
\end{equation}
Noncommutative effects are therefore manifest not only via $S$, which is
functionally different from its commutative quantum analog, but also directly
in the equation of motion for the canonical variables. This entails a series
of consequences for the model.\ The first of them is the condition for the
classical limit, which is now that the terms containing $Q_{K}$ and $Q_{I}$ be
negligible in (\ref{67}) and (\ref{69}). The presence of the $V_{nc}$
and$\ Q_{I}$\ terms in (\ref{69})\ tells us also\ that\ noncommutativity can
induce dynamics in situations where it is impossible in the commutative case.
Real wavefunctions ($S=0$), which represent universes that are necessarily
static in the commutative formulation,\footnote{Real wavefunctions are
priviledged, e.g., by the no-boundary proposal for the initial conditions of
the universe \cite{23}.} can yield dynamic universes in noncommutative quantum cosmology.

In what follows we present examples of application of the formalism proposed.

\subsection{Case 1}

The wavefunction is of the type (\ref{65}). In this case we have $S=\nu
\sqrt{3}\beta$. The equations of motion are therefore
\begin{equation}
\dot{\Omega}_{c}=0,\text{ \ \ \ \ }\dot{\beta}_{c}=2\sqrt{3}\nu-48\sqrt
{3}\theta\exp\left(  -2\sqrt{3}\Omega_{c}+3\theta\nu\right)  \label{77}%
\end{equation}
whose solution is
\begin{equation}
\Omega_{c}=\Omega_{c_{0}},\text{ \ \ \ }\beta_{c}(t)=2\sqrt{3}\alpha_{\nu
}\left(  t-t_{0}\right)  +\beta_{0}, \label{78}%
\end{equation}
where
\begin{equation}
\alpha_{\nu}=\left[  \nu-24\theta\exp\left(  -2\sqrt{3}\Omega_{c_{0}}%
+3\theta\nu\right)  \right]  . \label{78.5}%
\end{equation}
From (\ref{63}) and (\ref{64}) we have
\begin{align}
\Omega(t)  &  =\Omega_{c_{0}}-\frac{\theta\sqrt{3}}{2}\nu\nonumber\\
\beta(t)  &  =\beta_{c}(t)=2\sqrt{3}\alpha_{\nu}\left(  t-t_{0}\right)
+\beta_{0} \label{79}%
\end{align}
Except by the $\theta$ contributions that appear shifting the values of the
constants, the time dependence of the solutions (\ref{79}) is exactly the same
as the commutative counterpart discussed\ in section 6.\ The qualitative
behavior assumed by the universe in this case are therefore identical to the
one discussed there.

\subsection{Case 2}

Let us now consider a wavefunction that is the combination of two solutions of
the type (\ref{65}),%

\begin{equation}
\Psi(\Omega_{c},\beta_{c})=A_{1}K_{i\mu}\left(  4e^{-\sqrt{3}\Omega
_{c}+3\theta\mu/2}\right)  e^{i\sqrt{3}\mu\beta_{c}}+A_{2}K_{i\nu}\left(
4e^{-\sqrt{3}\Omega_{c}+3\theta\nu/2}\right)  e^{i\sqrt{3}\nu\beta_{c}}.
\label{90}%
\end{equation}
By writing it in the polar form we can find its phase as%

\begin{equation}
S(\Omega_{c},\beta_{c})=\arctan\left[  \frac{A_{1}K_{i\mu}\left(
4e^{-\sqrt{3}\Omega_{c}+3\theta\mu/2}\right)  \sin\left(  \sqrt{3}\mu\beta
_{c}\right)  +A_{2}K_{i\nu}\left(  4e^{-\sqrt{3}\Omega_{c}+3\theta\nu
/2}\right)  \sin\left(  \sqrt{3}\nu\beta_{c}\right)  }{A_{1}K_{i\mu}\left(
4e^{-\sqrt{3}\Omega_{c}+3\theta\mu/2}\right)  \cos\left(  \sqrt{3}\mu\beta
_{c}\right)  +A_{2}K_{i\nu}\left(  4e^{-\sqrt{3}\Omega_{c}+3\theta\nu
/2}\right)  \cos\left(  \sqrt{3}\nu\beta_{c}\right)  }\right]  , \label{91}%
\end{equation}
where the $A_{1}$ and $A_{2}$\ are chosen\ as real coefficients. The equations
of motion (\ref{61}) and (\ref{62}) for this state are
\begin{align}
\frac{d\Omega_{c}}{dt}  &  =8\sqrt{3}\text{ }\frac{A_{1}A_{2}\left[  K_{i\mu
}^{\prime}K_{i\nu}-K_{i\mu}K_{i\nu}^{\prime}\right]  e^{-\sqrt{3}\Omega_{c}%
}\sin\left[  \sqrt{3}\left(  \mu-\nu\right)  \beta_{c}\right]  }{\left(
A_{1}K_{i\mu}\right)  ^{2}+\left(  A_{2}K_{i\nu}\right)  ^{2}+2A_{1}%
A_{2}K_{i\mu}K_{i\nu}\cos\left[  \sqrt{3}\left(  \mu-\nu\right)  \beta
_{c}\right]  },\nonumber\\
& \nonumber\\
\frac{d\beta_{c}}{dt}  &  =2\sqrt{3}\text{ }\frac{\mu A_{1}^{2}K_{i\mu}%
^{2}+\nu A_{2}^{2}K_{i\nu}^{2}+\left(  \mu+\nu\right)  A_{1}A_{2}K_{i\mu
}K_{i\nu}\cos\left[  \sqrt{3}\left(  \mu-\nu\right)  \beta_{c}\right]
}{\left(  A_{1}K_{i\mu}\right)  ^{2}+\left(  A_{2}K_{i\nu}\right)  ^{2}%
+2A_{1}A_{2}K_{i\mu}K_{i\nu}\cos\left[  \sqrt{3}\left(  \mu-\nu\right)
\beta_{c}\right]  }\label{92}\\
& \nonumber\\
&  -48\sqrt{3}\theta e^{-2\sqrt{3}\Omega_{c}}\text{ }\frac{e^{3\mu\theta}%
A_{1}^{2}K_{i\mu}^{2}+e^{3\nu\theta}A_{2}^{2}K_{i\nu}^{2}+\left(
e^{3\mu\theta}+e^{3\nu\theta}\right)  A_{1}A_{2}K_{i\mu}K_{i\nu}\cos\left[
\sqrt{3}\left(  \mu-\nu\right)  \beta_{c}\right]  }{\left(  A_{1}K_{i\mu
}\right)  ^{2}+\left(  A_{2}K_{i\nu}\right)  ^{2}+2A_{1}A_{2}K_{i\mu}K_{i\nu
}\cos\left[  \sqrt{3}\left(  \mu-\nu\right)  \beta_{c}\right]  },\nonumber
\end{align}
where prime means derivative with respect to the argument. As a reference for
comparison with the commutative analog, let us fix $A_{1}=A_{2}=1/\sqrt{2}$
and\ consider first the case where $\mu=1/10$ and $\nu=-1/5.$ Figure $4$
presents a plot of the velocity field associated with the differential
equations (\ref{92}) for $\theta=-4.$ The field of velocities suggests that
the ensemble of solutions in this case is similar as the one of the
commutative counterpart. When comparing each individual commutative solution
with its noncommutative analog we expect therefore to find it quantitatively
corrected. If qualitatively different, it should assume a behavior similar to
one of the previously described in section 6. We shall verify this by studying
the evolution of $l^{3}(t)$.

Fig. $5a$ presents a plot of $\ln\left[  l^{3}(t)\right]  $ for\ a cyclic
universe where $\theta=4$ and the initial conditions identical\footnote{In all
cases of Fig. 5 initial conditions for $\Omega_{c}$ and $\beta_{c}$ were
chosen judiciously in order that the associated $\Omega\left(  0\right)  $ and
$\beta\left(  0\right)  $ calculated using (\ref{63}) and (\ref{64})
correspond to representative examples of the noncommutative quantum dynamics.}
to that of Fig. $3a$. As it can be seen, the principal effect of
noncommutativity in this case is to shorten the period of the cycles. In a
similar way as in the classical noncommutative case, the noncommutative
effects are sensible to the $\theta$ sign. Fig. $5b$ presents the plot of the
solution obtained by preserving the initial conditions of Fig. $5a$ and
inverting the sign of $\theta.$ The orbit which was originally closed
(corresponding to a non-singular universe) now\ becomes open, originating a
singular contracting universe. Note that the RHS of the system (\ref{92}) has
its sign inverted by the change $\mu\rightarrow-\mu$, $\nu\rightarrow-\nu$ and
$\theta\rightarrow-\theta$. By differentiating equations (\ref{63}) and
(\ref{64}) and using equations (\ref{91}) and (\ref{92}), we can see that
$\dot{\Omega}(t)$ and $\dot{\beta}(t)$ have\ their signs inverted by the same
change. It is therefore possible to generate an expanding universe solution
from the solution depicted\ in Fig. $5b$ by using this property.
Noncommutativity\ can also close orbits which were originally open. An example
of this property is depicted in Fig. $5c$, which present a cyclic universe
whose correspondent commutative analog is the universe solution of Fig. $3b$.
For large values of $\theta$ the effect of closing the orbit can be reverted,
as is shown in Fig. $5d.$%

\begin{center}
\includegraphics[
trim=0.000000in 0.000000in -0.155362in -0.168420in,
height=3.563in,
width=4.4166in
]%
{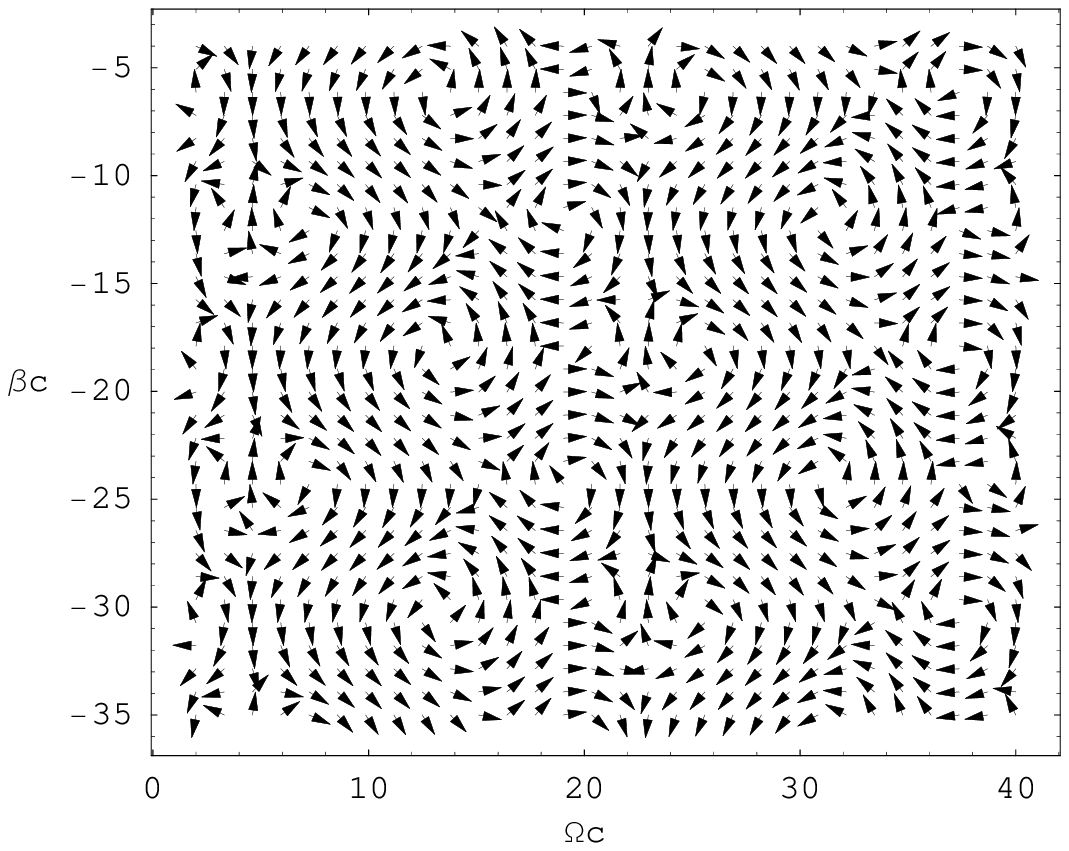}%
\\
FIG. 4. The normalized field of velocities corresponding to the Bohmian
differential equations for noncommutative Kantowski-Sachs universe with
$\mu=1/10,\nu=-1/5$ and $\theta=4.$%
\end{center}
We close this section by considering the case where $\mu=-\nu.$ As discussed
before, although this case has a real wavefunction it can have a nontrivial
dynamics. In fact, it can be shown that, depending on the initial conditions,
it can give rise to non-singular periodic solutions similar to that of Fig.
$5a$ or to singular universes similar to that of Fig. $5b.$\ %

\begin{center}
\includegraphics[
height=4.0352in,
width=5.9482in
]%
{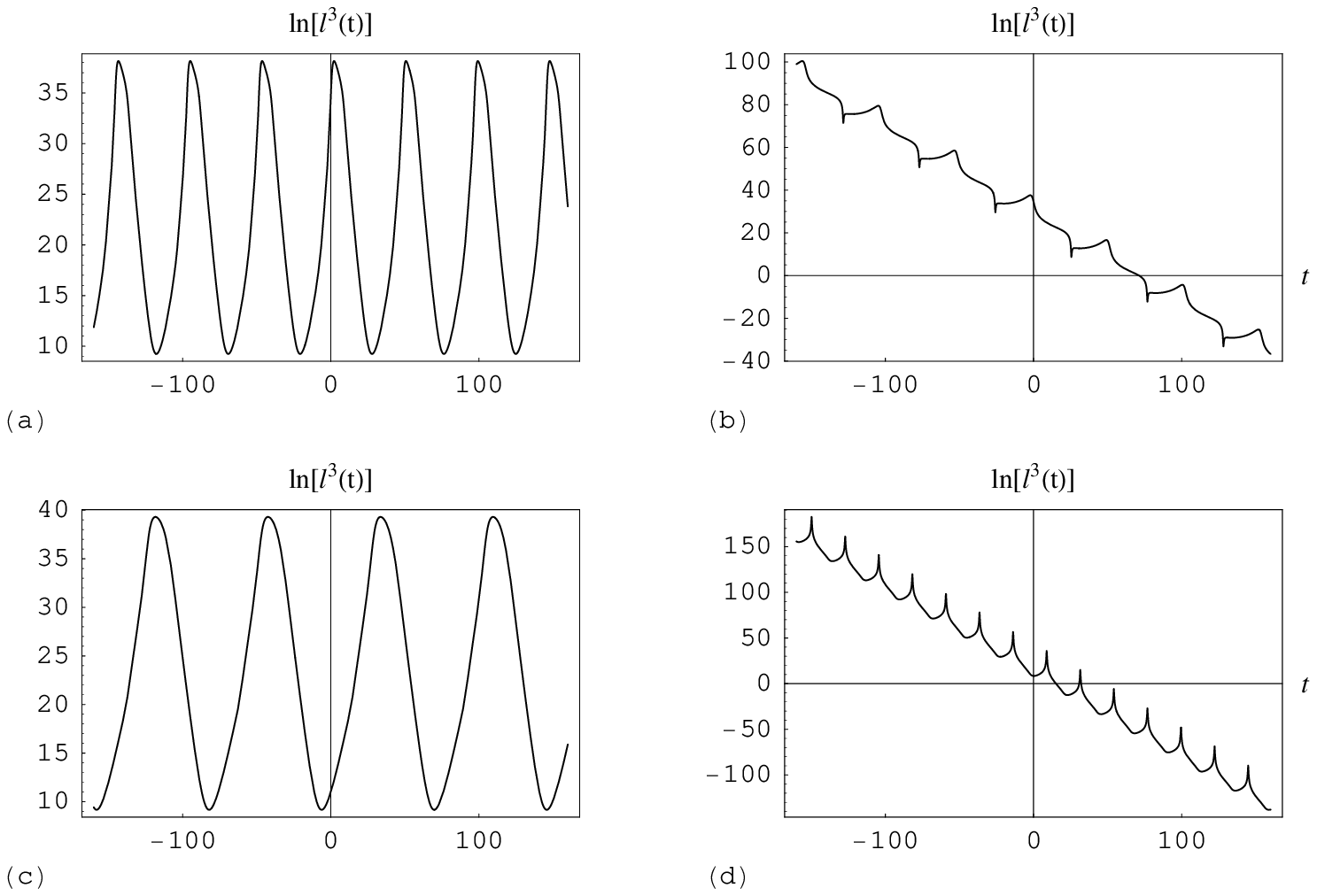}%
\\
FIG. 5. The evolution of the characteristic volume of the noncommutative
quantum Kantowski-Sachs universe $\left(  a\right)  :$ $\mu=1/10$, $\nu=-1/5$,
$\theta=4$ and $\Omega(0)=2,$ $\beta(0)=-24.$ $\ \left(  b\right)  :$
$\mu=1/10$, $\nu=-1/5$, $\theta=-4$ and $\Omega(0)=2,$ $\beta(0)=-24$.
$\ \left(  c\right)  :$ $\mu=1/10$, $\nu=-1/5$, $\theta=1$ and $\Omega
(0)=7.7,$ $\beta(0)=-21.7.$ $\ \left(  d\right)  :$ $\mu=1/10$, $\nu=-1/5$,
$\theta=12$ and $\Omega(0)=7.7,$ $\beta(0)=-21.7.$%
\end{center}

\section{Discussion and Outlook}

In this work we investigated some possible\ consequences of noncommutativity
for cosmology\ in the early stages of the universe\ via deformation of the
commutation relation between the minisuperspace variables. As an object of
analysis, we chose the Kantowski-Sachs universe, which is one of the most
investigated homogeneous universe models. In order to have a clear picture of
the impact of noncommutativity, a comparative study of the Kantowski-Sachs
universe was carried out in four different scenarios: the classical
commutative, classical noncommutative, quantum commutative and quantum
noncommutative. Our comparative analysis of the four versions of the model was
traced out in the common language of minisuperspace\ trajectories. In the
quantum context, this is provided by the Bohmian interpretation. We were lead,
therefore, to\ combine noncommutative geometry and Bohmian quantum physics.
This fusion of two apparently opposite ways of thinking, one commonly
associated with\ fuzzyness, and the other to ontological point particles,
proved to be interesting from the conceptual as well as from the operational
point of view.

In the theoretical framework, we have extended\ Bohmian formulation to
comprise the noncommutativity of the minisuperspace operators of the
Kantowski-Sachs model via the beable correspondence. Such a mapping between
Hermitian operators and ordinary functions, commonly employed in Bohmian
quantum mechanics, allows the association of each Hermitian operator with an
element of ontology. In that context, it can be shown that by averaging the
beable $\mathcal{B[}\hat{A}]$ over an ensemble of particles with probability
density $\rho=\left|  \Psi\right|  ^{2}$ gives the same result obtained by
computing the expectation value of the observable $\hat{A}$ applying the
standard operatorial formalism, reason for the denomination ``local
expectation value'' for $\mathcal{B[}\hat{A}]$ \cite{20}.\ In the
Kantowski-Sachs universe, although the Wheeler-DeWitt equation is of
Klein-Gordon type (it gives us no natural notion of probability), the beable
mapping is well defined, even in the noncommutative case. In the commutative
context, our formulation is reduced to the Bohmian quantum gravity proposed by
Holland \cite{20} in the minisuperspace approximation.

From the practical point of view, the formulation proposed proved to be easy
to handle in the calculations. The worked examples showed that
noncommutativity in the classical context (section 5) can modify
quantitatively the universe volume evolution and\ qualitatively its shape at
intermediary times, but cannot alter its singular behavior in the infinite
future and in the past. Quantum effects, on the other hand, can\ radically
modify the universe evolution and remove singularities. Some
quantum\ universes start with infinite volume and decrease up to a singular
configuration or start from a singularity in past and increase in volume up to
infinity. More interesting are the quantum periodic solutions, where the
singularities are totally removed. As showed in section 6, these universes can
present a great number of $e$-folds. The minimum length, $l_{\min}$, for these
eternal universes can assume a wide range of values. It is not difficult to
find solutions in which $l_{\min}$ is sufficiently small to be in a scale
where quantum gravity effects are expected to be relevant, but larger than the
Planck length, where a fundamental theory of gravitation is expected to be valid.

In section 7 it was shown that noncommutativity can modify appreciably the
universe evolution in the quantum context.\ A comparison between the two
quantum versions of the Kantowski-Sachs universe revealed that noncommutative
effects can not only introduce quantitative corrections in the universe
evolution but also modify its qualitative behavior. Periodic solutions can
change to exponentially contracting\ or expanding universes, and vice-versa.
The presence of noncommutative terms in the Bohmian equation of motion
(\ref{69}) for $\beta_{c}(t)$ tells us that noncommutativity can give rise to
a nontrivial dynamics when the wavefunction is real.

Although the analysis carried out in this work was restricted to the
Kantowski-Sachs model, part of the results may be valid to other homogeneous
cosmologies. Since the Friedman-Robertson-Walker universe presents a
Wheeler-DeWitt equation similar to the one discussed here (see, e.g.,
\cite{23.5}) the formalism proposed in this work may also be applied in its description.

\section*{Acknowledgments}

The authors acknowledge Leonardo Guimar\~{a}es de Assis for his useful help in
the preparation of the figures. This work was financially supported by CAPES
and CNPq.

\end{document}